\documentclass[12pt]{amsart}
\usepackage{graphicx}
\usepackage{epsfig}
\usepackage{amsmath}
\usepackage{amsxtra}
\usepackage{amscd}
\usepackage{amsfonts}
\usepackage{amssymb}
\usepackage{eucal}

\newcommand{\C}{{\mathbb C}}
\newcommand{\Z}{{\mathbb Z}}

\newtheorem{thm}{Theorem}[section]

\numberwithin{equation}{section}

\begin{document}

\title{INTEGRABLE COMBINATORICS}

\author{P. Di Francesco}

\address{Institut de Physique Th\'eorique, CEA Saclay IPHT,
Unit\'e de Recherche Associ\'ee du CNRS,
91191 Gif sur Yvette Cedex, France.
E-mail: philippe.di-francesco@cea.fr}

%

\begin{abstract}
We review various combinatorial problems with underlying classical or quantum integrable structures.
(Plenary talk given at the International Congress of Mathematical Physics, Aalborg, 
Denmark, August 10, 2012.)
\end{abstract}

\keywords{integrability, quantum gravity, integrable lattice model, cluster algebra, 
Laurent phenomenon, tree, alternating sign matrix, plane partition, lattice paths, networks.}
\maketitle

\section{Introduction}

In these notes, we present a few mathematical problems or constructs, most of them combinatorial in nature, that
either were introduced to explicitly solve or better understand physical questions (Mathematical Physics)
or can be better understood in the light of physical interpretations (Physical Mathematics). The frontier
between the two is subtle, and we will try to make this more concrete in the examples chosen below.

We start from two different physical theories of discretized random surfaces, describing the possible fluctuations
of an underlying two-dimensional discrete space time or equivalently of a discretized 
Lorentzian\cite{AL} (Sect. 2) or Euclidian\cite{DISC} (Sect. 3) 1+1- or 2-dimensional metric. 
In both simple models, we show the existence
of some hidden integrable structure, in two very different forms. The first model may be equipped with an infinite
family of commuting transfer matrices\cite{LORGRA} governing the time-evolution of the surfaces, thus displaying quantum 
integrability, with an infinite number of conserved quantities. The second model actually addresses
the correlations of marked points on the discrete surface at prescribed geodesic distance, and we show that such correlations
viewed as evolutions in the geodesic distance variable,
form a discrete classical integrable system with some conserved quantities modulo the equations of motion\cite{GEOD}.
Remarkably, both cases can be reduced bijectively to statistical ensembles of trees.

Few combinatorial objects form such a vivid crossroads between various physical and mathematical theories and constructs
as the Alternating Sign Matrices\cite{Bressoud}. These were defined by Robbins and Rumsey\cite{RR} in an attempt to generalize the notion
of determinant of a square matrix, while keeping certain properties, in particular the Laurent polynomiality of the
result in terms of the matrix elements. Interestingly, this happened two decades before a combinatorial theory
of the Laurent phenomenon was discovered by Fomin and Zelevinsky, under the name of Cluster Algebra\cite{FZI}. 
In Sect. 4, we try to unravel the thread from alternating sign matrices to integrable lattice models, back to more combinatorial
objects such as plane partitions. We show that the refined enumeration of such objects involves underlying integrable
structures similar to that of discrete Lorentzian  gravity. This fact is actually used to prove the 
Mills Robbins Rumsey conjecture\cite{MRR}
relating alternating sign matrices to descending plane partitions.
Section 5 is devoted to the cluster algebra formulation and explicit solution
of the celebrated T-system\cite{KNS}, a discrete 2+1-dimensional integrable system\cite{KLWZ} related
to alternating sign matrices, but also to domino tilings of the Aztec diamond\cite{EKLP,SPY} and 
Littlewood-Richardson coefficients\cite{KTW}. 
We show that solving such a system amounts to computing partition functions of weighted path
models, a kind of discrete path integral, that gives a new insight on the Laurent positivity conjecture of cluster algebra.

\section{1+1D Lorentzian gravity, integrability and trees}\label{lorsec}

\subsection{1+1D Lorentzian gravity}
Discrete models for 1+1D Lorentzian gravity are defined as follows.  They involve
a statistical ensemble of discrete space-times, which may be modeled by random triangulations
with a regular time direction (a segment $[t_1,t_2]$) and a random space direction, obtained
by random triangulations of unit time strips $[t,t+1]$ by arbitrary but finite numbers of triangles with
one edge along the time line $t$ (resp. $t+1$) and the opposite vertex on the time line $t+1$ (resp. $t$).
All other edges are then glued to their neighbors so as to form a triangulation. The boundary may be taken
free, periodic or staircase-like on the left \cite{LORGRA}.
A typical such Lorentzian triangulation $\Theta$  in 1+1D reads as follows:
$$ \raisebox{-1.cm}{\hbox{\epsfxsize=8.cm \epsfbox{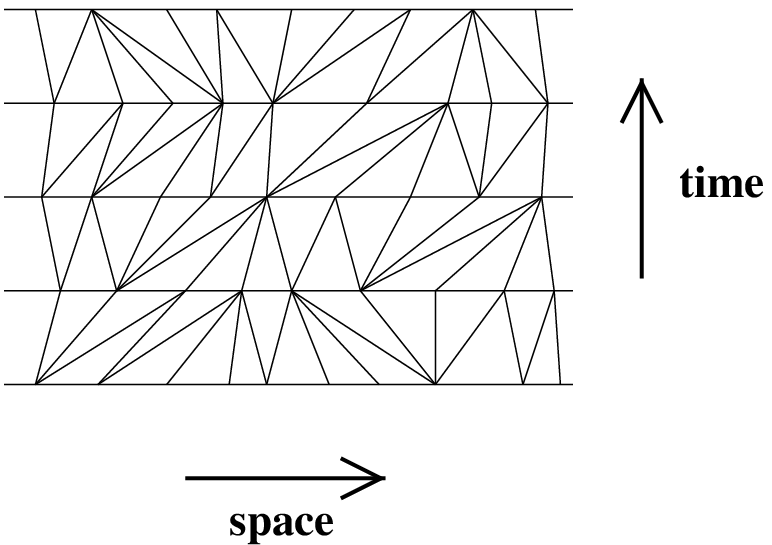}}}  $$
These triangulations are best described in the dual picture by considering triangles as vertical half-edges
and pairs of triangles that share a time-like (horizontal) edge as vertical edges between two consecutive time-slices.
We may now concentrate on the transition between two consecutive time-slices which typically reads as follows:
\begin{equation}\label{transmatbare} \raisebox{-1.cm}{\hbox{\epsfxsize=8.cm \epsfbox{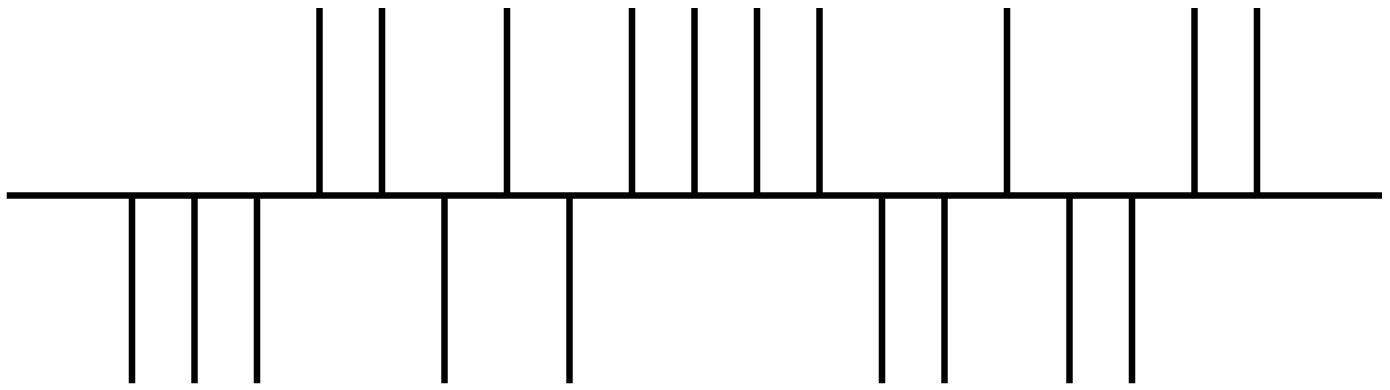}}}  \end{equation}
with say $i$ half-edges on the bottom and $j$ on the top (here for instance we have $i=9$ and $j=10$). 
Denoting by $\vert i\rangle$ and $\vert j\rangle$, 
$i,j\in \Z$ the bottom and top states bases, we may describe the generation of a triangulation 
by the iterated action of a transfer operator ${\mathcal T}$ with matrix elements $T_{i,j}={i+j\choose i}$.
Note that the corresponding matrix $T=(T_{i,j})_{i,j\in \Z_+}$ is infinite. We shall deal with such matrices in the following.
A useful way of thinking of them is via the double generating function:
$$ f_T(z,w)=\sum_{i,j\geq 0} T_{i,j} z^i w^j =\frac{1}{1-z-w} \, .$$

\subsection{Integrability}

To make the model more realistic, we may include both area and curvature-dependent terms, by introducing
Botlzmann weights $w(\Theta)$ equal to the product of local weights of the form $g$ per triangle (area term)
and $a$ per pair of consecutive triangles in a time-slice pointing in the same direction (both up or both down).
The rules in the dual picture are as follows:
$$\raisebox{-1.cm}{\hbox{\epsfxsize=6.cm \epsfbox{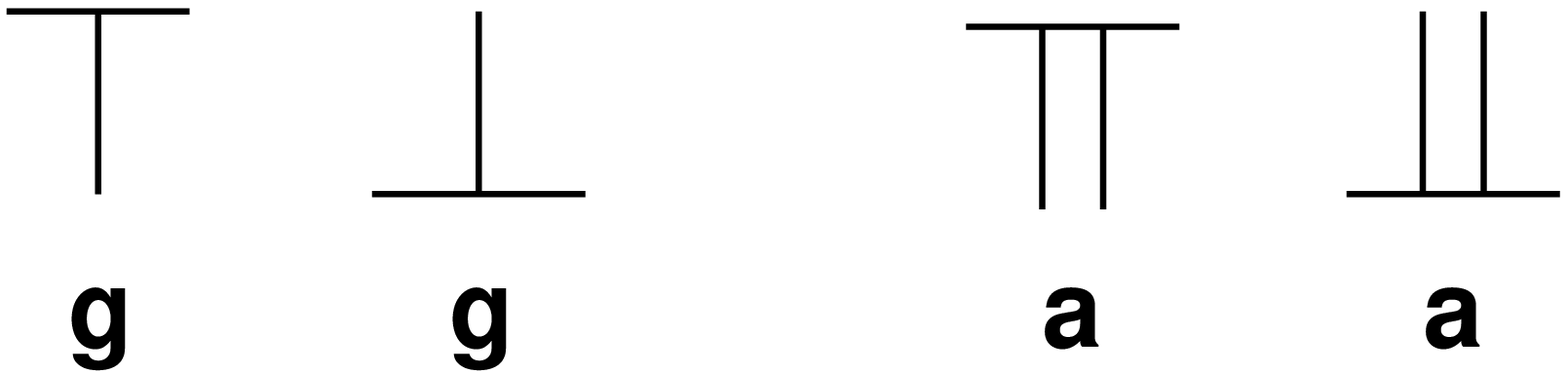}}}   $$
For instance, in the example \eqref{transmatbare} above with $i=9$ and $j=10$, the product of local weights is $g^{19} a^{9}$.
It is easy to see that these weights correspond to a new transfer operator ${\mathcal T}(g,a)$ with
matrix elements 
$$T(g,a)_{i,j}= (ag)^{i+j}\, \sum_{k=0}^{{\rm Min}(iij)} {i\choose k}{j\choose k} a^{-2k} $$
Equivalently, the double generating function reads:
\begin{equation}\label{simple} f_{T(g,a)}(z,w)=\sum_{i,j\geq 0} T(g,a)_{i,j} z^i w^j =\frac{1}{1-ga(z+w)-g^2(1-a^2)z w} 
\end{equation}

This model turns out to provide one of the simplest examples of quantum integrable system, with an infinite
family of commuting transfer matrices. Indeed, we have:

\begin{thm}\cite{LORGRA}
The transfer matrices $T(g,a)$ and $T(g',a')$ commute if and only if the parameters $(g,a,g',a')$ are such that
$\varphi(g,a)=\varphi(g',a')$ where:
$$ \varphi(g,a)=\frac{1-g^2(1-a^2)}{a g} $$
\end{thm}

This is easily proved by using the generating functions, and noting that for two infinite matrices $A,B$
we have formally $f_{AB}=f_A\star f_B$, where $\star$ stands for the convolution product, namely
$$ (f_A \star f_B) (z,w)=\oint_{\mathcal C} {dt \over 2i\pi t} f_A\left(z,\frac{1}{t}\right)\, f_B(t, w) $$
where the contour integral picks up the constant term in $t$.

This was extensively used\cite{LORGRA} to diagonalize ${\mathcal T}(g,a)$ and to compute correlation
functions of boundaries in random Lorentzian triangulations. Our purpose here was simply to display the very
simple form of the generating function \eqref{simple}, which will reappear later in these notes.

\subsection{Trees}

For suitable choices of boundary conditions, the dual random Lorentzian triangulations introduced above may
be viewed as random plane trees. This is easily realized by gluing all the bottom vertices of parallel vertical edges
whose both top and bottom halves contribute to the curvature term (no interlacing with the neighboring time slices).
A typical such example reads:
$$ \raisebox{-1.cm}{\hbox{\epsfxsize=10.cm \epsfbox{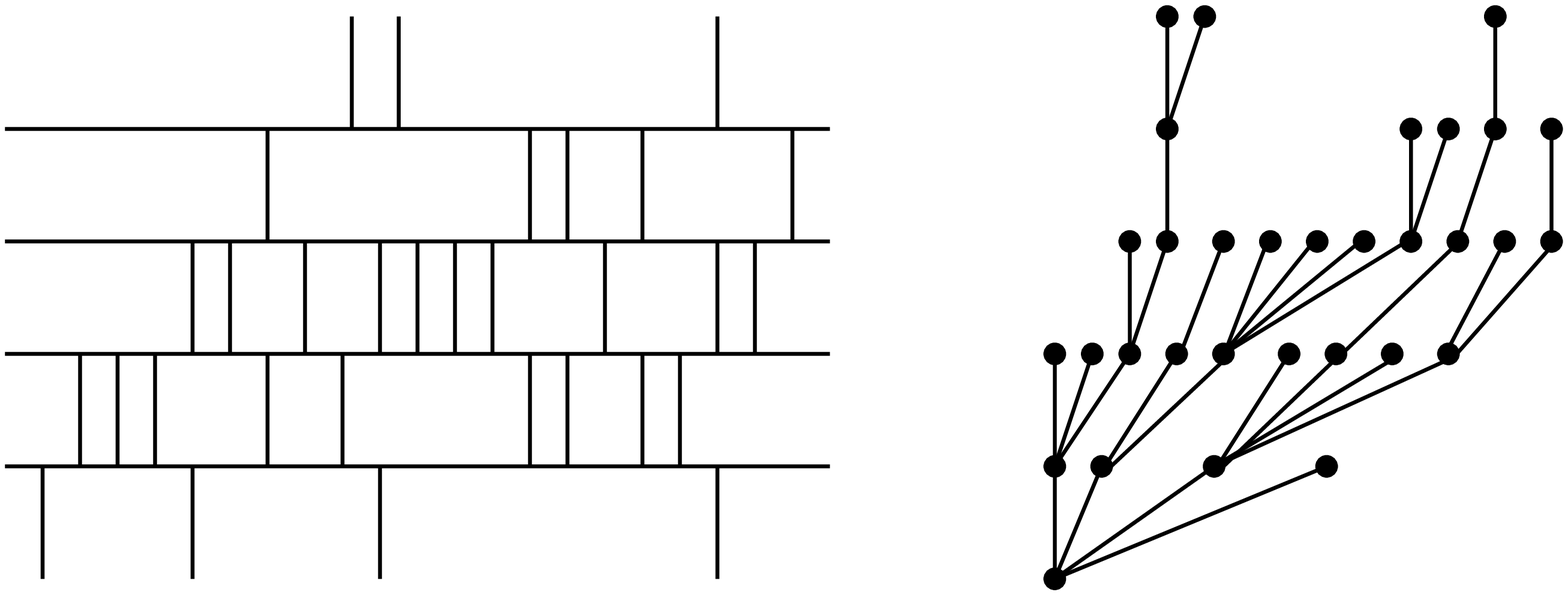}}}$$
Note that the tree is naturally rooted at its bottom vertex.

To summarize, we have unearthed some integrable structure attached naturally to plane trees, one of the most fundamental
objects of combinatorics. Note that in tree language the weights are respectively $g^2$ per edge, and $a$ 
per pair of consecutive descendent edges and per pair of consecutive leaves at each vertex (from left to right).

\section{Planar maps and geodesics}

\subsection{Two-dimensional quantum gravity}

Two-dimensional quantum gravity (2DQG) is a theory describing the interactions of matter with the underlying
space-time, in which both are quantized. Discrete models of 2DQG involve statistical matter models defined
on statistical ensembles of discrete random surfaces, in the form of random tessellations. The Einstein action
for these 2D random surfaces involves their two invariants, the area and the genus. The complete model will therefore
combine the Boltzmann weights of the statistical matter model and of the underlying surface, in the form of 
parameters coupled to its area and its genus. Matrix integrals have proven to be extremely powerful tools for
generating such discrete random surfaces with matter, by noting that the expansion for large matrix size 
matches the genus expansion. In a parallel way,
the field theoretical descriptions of the (critical) continuum limit of 2DQG have blossomed into a
more complete picture with identification of relevant operators and computation of their correlation functions.
This was finally completed by an understanding in terms of the intersection theory of the moduli space of curves with punctures
and fixed genus. Remarkably, in all these approaches a common integrable structure is always present. It takes
the form of commuting flows in parameter space.

However, a number of issues were left unadressed by the matrix/field theoretical approaches. What about the intrinsic
geometry of the random surfaces? Correlators must be integrated w.r.t. the position of their insertions, leaving us
only with topological invariants of the surfaces. But how to keep track say of the geodesic distances between two
insertion points, while at the same time summing over all surface fluctuations?

\subsection{Maps and trees}

Answers to these questions came from a better combinatorial understanding of the structure of the (planar) tessellations 
involved in the discrete models. And, surprisingly, yet another form of integrability appeared. Following pioneering
work of Schaeffer \cite{SCH}, it was observed that all models of discrete 2DQG with a matrix model solution 
(at least in genus 0) could be expressed as statistical models of (decorated) trees, and moreover, the decorations
allowed to keep track of geodesic distances between some faces of the tessellations. Marked planar tessellations of
physicists are known as rooted planar maps in mathematics. They correspond to connected graphs (with vertices, edges, faces)
embedded into the Riemann sphere. Such maps are usually represented on a plane with a distinguished face 
``at infinity", and a marked edge adjacent to that face. The degree of a vertex is the number of distinct half-edges
adjacent to it, the degree of a face is the number of edges forming its boundary.

Let us concentrate on the following example of tetravalent (degree 4) planar maps with 2 univalent (degree 1) vertices, 
one of which is singled out as the root.
The Schaeffer bijection associates to each of these a unique rooted tetravalent 
(with inner vertices of degree 4) tree called blossom-tree, with two types
of leaves (black and white), and such that there is exactly one black leaf attached to each inner vertex.
$$  \raisebox{-1.cm}{\hbox{\epsfxsize=10.cm \epsfbox{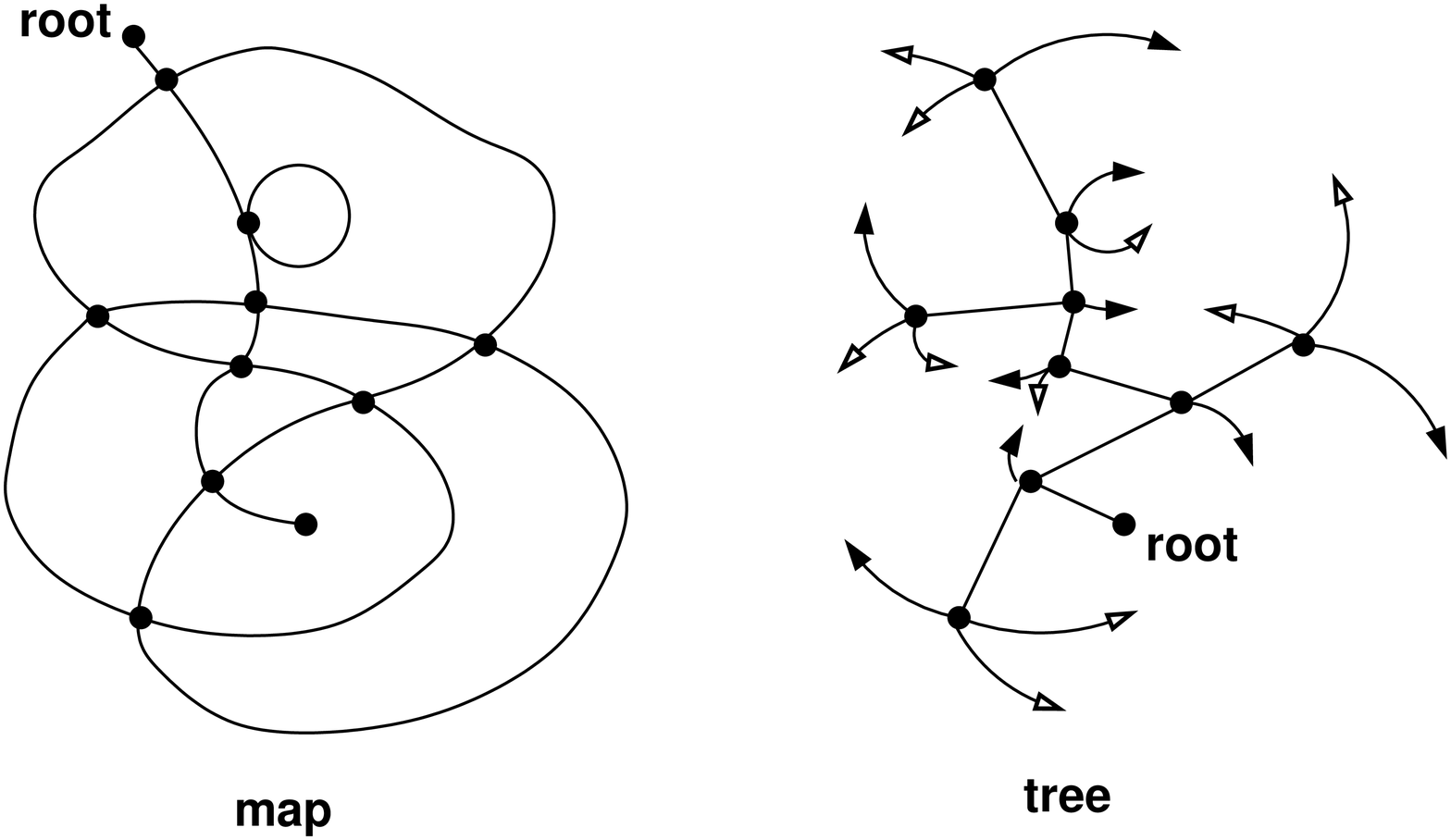}}}$$
This is obtained by the following cutting algorithm: travel clockwise along the bordering edges of the face at infinity, starting from
the root. 
For each traversed edge, cut it if and only if after the cut, the new graph remains connected, 
and replace the two newly formed half-edges
by a black and a white leaf respectively in clockwise order. Once the loop is traveled, this has created a larger face at infinity.
Repeat the procedure until the graph has only one face left: it is the desired blossom-tree, which we reroot at the other
univalent vertex, while the original root is transformed into a white leaf. 

Why is it a bijection? Remarkably, the information of black and white leaves is sufficient to close up the blossom-tree 
into a unique map. Clockwise around the tree, match each black leaf to its immediate follower if it is white, and repeat until
all but one leaves are matched. The loser of the musical chairs is the root of the map.

\subsection{Exact enumeration and integrability}

The bijection above allows to keep track of the geodesic distance between the faces containing the two univalent vertices,
defined as the minimal number of edges crossed in paths between the two faces
(this distance is 2 in the example). Let us introduce a weighted counting of the maps, by including a weight $g$ per vertex.

Defining $R_n(g)$ to be the generating function for maps with geodesic distance $\leq n$ 
between the two univalent vertices, we have the following relation:
\begin{equation}\label{crucial}
R_n(g)= 1+ gR_n(g)\left( R_{n+1}(g)+R_n(g)+R_{n-1}(g)\right) 
\end{equation}
easily derived by inspecting the environment of the vertex attached to the root of the tree when it exists.

This recursion relation must be supplemented with boundary conditions. First, when $n=0$, the relation \eqref{crucial}
makes sense only if the term $R_{-1}(g)$ is omitted, so we set $R_{-1}(g)=0$. Moreover, if we take $n\to \infty$,
we simply relax the distance condition, and the function $R(g)=\lim_{n\to \infty} R_n(g)$ is the generating function
for maps with two univalent vertices. It satisfies the limiting equation $R=1+3 g R^2$, and we easily get 
\begin{equation}\label{Rval}R(g) =\frac{1-\sqrt{1-12g}}{6 g} \end{equation}
as the unique solution with a formal power series expansion of the form $R(g)=1+O(g)$.

The function $R_n(g)$ is the unique solution to \eqref{crucial} for $n\geq 0$ such that 
(i) $R_{-1}(g)=0$ and (ii) $\lim_{n\to \infty}R_n(g)=R(g)$ of \eqref{Rval}. A first remark is in order:
the equation \eqref{crucial}, viewed as governing the evolution of the quantity $R_n(g)$ in the
discrete time variable $n$,  is a classical {\it discrete integrable system}. By this we mean that it has a
{\it discrete integral of motion}, expressed as follows. The function $\phi(x,y)$ defined by
\begin{equation}\label{cons}\phi(x,y)= x y (1-g(x+y))-x-y \end{equation}
is such that for any solution $S_n$ of the recursion relation \eqref{crucial}, the quantity 
$\phi(S_n,S_{n+1})$ is independent of $n$. In other words, the quantity  $\phi(S_n,S_{n+1})$ is 
conserved modulo \eqref{crucial}. This is easily shown by factoring $\phi(S_n,S_{n+1})-\phi(S_{n-1},S_n)$.

By the conservation of $\phi(R_n(g),R_{n+1}(g))$, we find that 
$$ \phi(R_n(g),R_{n+1}(g))=\lim_{m\to\infty} \phi(R_m(g),R_{m+1}(g))=\phi(R(g),R(g))$$
which gives us an explicit relation between $R_n(g)$ and $R_{n+1}(g)$. It turns out that we can
solve explicitly for $R_n(g)$:

\begin{thm}\cite{GEOD}\label{geodthm}
The generating function $R_n(g)$ for rooted tetravalent planar maps with two univalent vertices
at geodesic distance at most $n$ from each other reads:
$$ R_n(g)=R(g) \frac{(1-x(g)^{n+1})(1-x(g)^{n+4})}{(1-x(g)^{n+2})(1-x(g)^{n+3})} $$
where $x(g)$ is the unique solution of the equation:
$$ x+\frac{1}{x}+4=\frac{1}{g R(g)^2} $$
with a power series expansion of the form $x(g)=g+O(g^2)$.
\end{thm}

The form of the solution in Theorem \ref{geodthm} is that of a discrete soliton with tau-function $\tau_n=1-x(g)^n$.
Imposing more general boundary conditions on the equation \eqref{crucial} leads to elliptic solutions of the same flavor.
The solution above and its generalizations to many classes of planar maps have allowed for a better understanding
of the critical behavior of surfaces and their intrinsic geometry. Recent developments include planar three-point
correlations, as well as higher genus results.

To summarize, we have seen yet another integrable structure emerge in relation to (decorated) trees.
This is of a completely different nature from the one discussed in Section \ref{lorsec}, 
where a quantum integrable structure
was attached to rooted planar trees. Here we have a discrete classical integrable system, with soliton-like solutions.
Such structures will reappear in the following, in relation to (fixed) lattice statistical models.

\section{Alternating sign matrices}

\subsection{Lambda-determinant and Alternating Sign Matrices}

The definition of the so-called Lambda-determinant of Robbins and Rumsey \cite{RR} is based on the famous
Dodgson condensation algorithm \cite{DOD} for computing determinants, itself based on the Desnanot-Jacobi
equation, a particular Pl\"ucker relation, relating minors of any square $k+1\times k+1$ matrix $M$:
\begin{equation}\label{desnajac}
\vert M\vert \times \vert M_{1,k+1}^{1,k+1}\vert=\vert M_{k+1}^{k+1}\vert \times \vert M_1^1\vert -
\vert M_1^{k+1}\vert \times \vert M_{k+1}^1\vert
\end{equation}
where $\vert M_{i_1,i_2,...,i_r}^{j_1,j_2,...,j_r}\vert$ stands for the determinant of the matrix obtained from $M$
by deleting rows $i_1,...,i_r$ and columns $j_1,...,j_r$. The relation \eqref{desnajac} may be used as a
recursion relation on the size of the matrix, allowing for efficiently compute its determinant.

More formally, we may recast the algorithm using the so-called $A_\infty$ $T$-system 
(also known as discrete Hirota) relation:
\begin{equation}\label{tsys}
T_{i,j,k+1}T_{i,j,k-1}=T_{i,j+1,k}T_{i,j-1,k}-T_{i+1,j,k}T_{i-1,j,k}
\end{equation}
for any $i,j,k\in \Z$ with fixed parity of $i+j+k$. 
Now let $A=(a_{i,j})_{i,j\in \{1,2,...,n\}}$ be a fixed $n\times n$ matrix. Together with the initial data:
\begin{eqnarray}
T_{\ell,m,0}&=&1\quad \qquad \qquad\qquad (\ell,m\in \Z;\ell+m=n\, {\rm mod}\, 2)\nonumber \\ 
T_{i,j,1}&=&a_{{j-i+n+1\over 2},{i+j+n+1\over 2}} \quad
(i,j\in\Z;i+j=n+1\, {\rm mod}\, 2; |i|+|j|\leq n-1)\, ,\label{initdat}
\end{eqnarray}
the solution of the $T$-system \eqref{tsys} satisfies:
\begin{equation}
T_{0,0,n}=\det(A)
\end{equation}

Given a fixed formal parameter $\lambda$,
the Lambda-determinant of the matrix $A$, denoted by $\vert A\vert_\lambda$ is simply defined 
as the solution $T_{0,0,n}=\vert A\vert_\lambda$ of the deformed $T$-system
\begin{equation}\label{defoT} 
T_{i,j,k+1}T_{i,j,k-1}=T_{i,j+1,k}T_{i,j-1,k}+\lambda \, T_{i+1,j,k}T_{i-1,j,k}
\end{equation}
subject to the initial condition \eqref{initdat}.

The discovery of Robbins and Rumsey is that the Lambda-determinant is a homogeneous
Laurent polynomial of the matrix entries of degree $n$, and that moreover the monomials
in the expression are coded by $n\times n$ matrices $B$ with entries $b_{i,j}\in \{0,1,-1\}$, characterized by the fact that
their row and column sums are $1$ and that the partial row and column sums are non-negative, namely
\begin{eqnarray*} \sum_{i=1}^k b_{i,j} \geq 0 & & \quad \sum_{i=1}^k b_{j,i}\geq 0\quad (k=1,2,...,n-1;j=1,2,...,n)\\
 \sum_{i=1}^n b_{i,j} = 1 & & \quad \sum_{i=1}^n b_{j,i}=1\quad (j=1,2,...,n)
\end{eqnarray*}
Such matrices $B$ are called alternating sign matrices (ASM). These include the permutation matrices (the ASMs
with no $-1$ entry). Here are the 7 ASMs of size 3:
$$ \begin{pmatrix}1 &0&0\\0&1&0\\0&0&1\end{pmatrix}\quad
\begin{pmatrix}0 &1&0\\1&0&0\\0&0&1\end{pmatrix}\quad
\begin{pmatrix}1 &0&0\\0&0&1\\0&1&0\end{pmatrix}\quad
\begin{pmatrix}0 &0&1\\0&1&0\\1&0&0\end{pmatrix}\quad
\begin{pmatrix}0 &1&0\\0&0&1\\1&0&0\end{pmatrix}\quad
\begin{pmatrix}0 &0&1\\1&0&0\\0&1&0\end{pmatrix}\quad
\begin{pmatrix}0&1&0\\1&-1&1\\0&1&0\end{pmatrix}$$
There is an explicit formula for the Lambda-determinant\cite{RR}:
\begin{equation}\label{lambdadet}
\vert A\vert_\lambda=\sum_{n\times n\, ASM\, B} \lambda^{{\rm Inv}(B)-N(B)} (1+\lambda)^{N(B)}\prod_{i,j} a_{i,j}^{b_{i,j}}
\end{equation}
where ${\rm Inv}(B)$ and $N(B)$ denote respectively the inversion number and the number of entries $-1$ in $B$,
with 
\begin{eqnarray*}
{\rm Inv}(B)&=&\sum_{1\leq i<j \leq n\atop 1\leq k<\ell \leq n} b_{i,\ell}b_{j,k} \\
N(B)&=&\frac{1}{2}\left( -n+\sum_{1\leq i,j\leq n} |b_{i,j}| \right) 
\end{eqnarray*}
Note that for $\lambda=-1$, only the ASMs with $N(B)=0$ contribute, i.e. the permutation matrices, for which
${\rm Inv}(B)$ coincides with the usual inversion number of the corresponding permutation, and therefore \eqref{lambdadet}
reduces to the usual formula for the determinant.

Robbins and Rumsey formulated the famous ASM conjecture, that the total number of $n\times n$ ASMs is
given by $A_n=\prod_{j=0}^{n-1} \frac{(3j+1)!}{(n+j)!}$.

\subsection{Integrabilities: six vertex, loop gas and more}

ASMs relate to two different kinds of integrable systems. 

On one hand, the $T$-system relation \eqref{tsys}
and its deformations \eqref{defoT}
are known to be discrete classical integrable systems\cite{KLWZ}, 
when interpreted as a 2+1D evolution in the discrete time variable $k$.
It is also part of a combinatorial structure called a cluster algebra\cite{FZI}, 
which we will discuss in the next section. As such, it has
a guaranteed Laurent property, namely that its solutions are always Laurent polynomials of the initial data, which explains
the miracle witnessed by Robbins and Rumsey.

On the other hand ASMs are in bijection with configurations of a famous
integrable 2D lattice model, the Six Vertex (6V) model with particular, so-called Domain Wall Boundary Conditions (DWBC)
\cite{KO,IZ}. 
This latter connection allowed Kuperberg to derive an elegant
proof of the ASM conjecture \cite{KUP}, shortly after Zeilberger's remarkable 
direct proof using generating functions\cite{ZEIL}.

\begin{figure}
\centering
\includegraphics[width=12.cm]{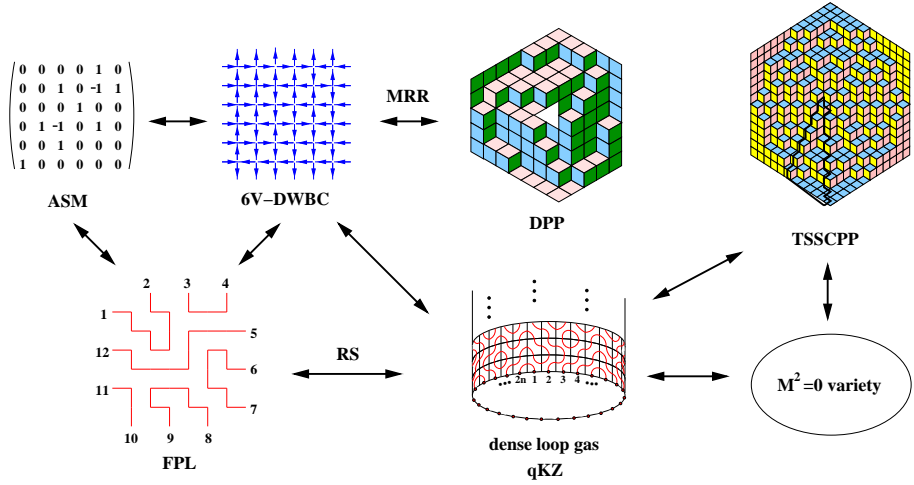}
\caption{\small The combinatorial family of ASMs. From left to right: ASM, 6V-DWBC and FPL , all in bijection;
dense loop gas: its groundstate/limiting probability vector satisfies the qKZ equation, the components measure 
FPL correlations (RS conjecture), their sum matches the 6V-DWBC partition function with inhomogeneous 
spectral parameters $z_i$ and $q^3=1$; DPP: their refined evaluation matches that of ASMs (MRR conjecture);
TSSCPP: their refined enumeration matches a sum rule for qKZ solutions at generic $q$ and $z_i=1$; Variety $M^2=0$: 
its degree/multidegree matches solutions of qKZ for $q=1$.}
\label{fig:everything}
\end{figure}

The connection with the 6V model is only the tip of an iceberg, which has kept revealing 
more structure over the years (see the sketch in Fig.\ref{fig:everything}).
The ASMs of size $n$ are indeed also in same number as: 

(i) the Totally Symmetric Self-Complementary Plane Partitions (TSSCPP), namely the rhombus tilings of a regular hexagon
of size $2n$ drawn on the triangular lattice, with all possible symmetries of the hexagon, which amount to the tilings of a 
fundamental domain occupying $\frac{1}{12}$ of the hexagon (see Fig.\ref{fig:everything}).

(ii) the Descending Plane Partitions (DPP) of order $n$, which we will define below.

There is however no known natural bijection between these objects, and it remains a challenge to find one.

The picture has grown more complex with the remark of Batchelor, DeGier and Nienhuis that the groundstate
of the so-called $O(n=1)$ dense loop model on a cylinder of perimeter $2n$ has positive rational entries summing up to $1$
with common denominator $A_n$. Alternatively, each component is interpreted as the limiting probability of loop configurations
realizing a fixed pattern of connection of the $2n$ string ends on the boundary of the cylinder.
Shortly after, the numerator of each component of this vector has been identified 
by Razumov and Stroganov (RS) as some particular correlation of the 6V model (and equivalently of ASMs), once reformulated
as a model of fully-packed loops (FPL) on a $n\times n$ square lattice grid \cite{RS}. 
This conjecture was proved recently by Cantini and Sportiello
\cite{CS} in a purely combinatorial way.

In the next subsections, we will address the connection between refined enumeration of ASMs and DPPs (the
Mills, Robbins and Rumsey (MRR) conjecture \cite{MRR}). Before we go into this, let us mention
that more connections have been found between the various combinatorial objects. The computation of the groundstate
of the dense loop gas with multi-parameter inhomogeneous Boltzmann weights was performed \cite{DZJ0} by use of the
so-called quantum Knizhnik-Zamolodchikov equation, whose suitably normalized solution yields the polynomial 
components of the groundstate vector, depending on $2n$ complex spectral parameters $z_i$, $i=1,2,...,2n$
and one quantum parameter $q$ parameterizing the weight per loop $-(q+q^{-1})$. it was shown \cite{DZJ0} that the sum
of these components equates the partition function of the 6V model with DWBC when $q$ is a non-trivial cubic root of unity,
expressed as a determinant by Izergin\cite{IZ}.
The ASM number is recovered in the limit when $z_i\to 1$ for all $i$.
On the other hand, when all $z_i=1$ but $q$ arbitrary, it was shown that the sum of the components of the groundstate
equates the partition function for TSSCPP on the fundamental domain above, with a particular weight $(q+q^{-1})$
for one type of tile \cite{DZJ1}. The solutions of the qKZ equation therefore form a bridge between ASM and TSSCPP, in that
two different specializations relate to either objects.

Finally, last but not least, yet another combinatorial problem turns out to be related to all of the above.
It was shown in \cite{DZJ2} that the degree
of the variety of strictly upper triangular $2n\times 2n$ complex matrices of vanishing square is equal to this 
weighted TSSCPP counting function, for $q=1$. Moreover, the variety actually decomposes into irreducible components
whose degrees are the entries of the vector, solution to the qKZ equation in the correlated limit where
$q\to 1$ and $z_i\to 1$.
More generally this extends to the  multi-degree of this variety, by considering its torus-equivariant cohomology, and by
taking a suitable limit of the qKZ solution \cite{DZJ2,KZ}. 

\subsection{ASMs from 6V}

The 6V-DWBC model on a square $n\times n$ grid has configurations obtained by choosing
an orientation for each edge of the square lattice, in such a way that the flow is conserved at each vertex,
namely exactly two edges are incoming and two outgoing at each vertex. Moreover the Domain Wall Boundary Conditions
impose that horizontal external edges to the grid point into the grid, while vertical ones point out of the grid.
This gives rise to the following 6 possible
local configurations:
$$\raisebox{-1.cm}{\hbox{\epsfxsize=12.cm \epsfbox{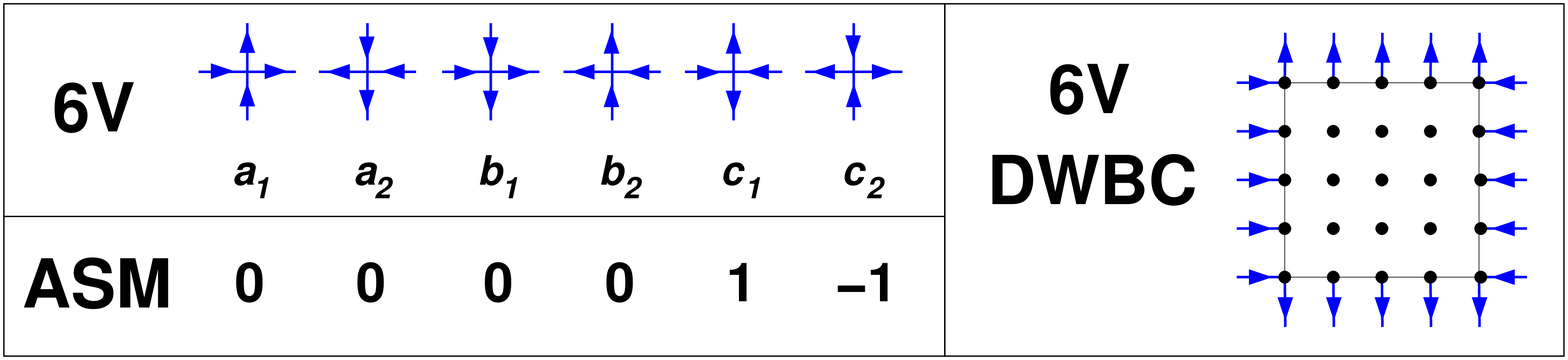}}} 
$$
where we also have indicated the corresponding ASM entries.
One usually attaches Boltzmann weights to each of the above configurations as follows: each row (resp. column)
of the grid carries a complex number $z_i$, $i=1,2,...n$ (resp. $w_i$, $i=1,2,...n$ ) called spectral parameter.
Moreover the weights depend on a ``quantum" parameter $q\in \C^*$.
We have the following parametrization of the weights:
$$ a(z,w)=q z-q^{-1} w \qquad b(z,w)=q^{-1} z -q w \qquad c(z,w)=(q^2-q^{-2}) \sqrt{z w} $$
where $a(z,w)$ is the weight for a vertex of type $a_1$ or $a_2$ at the intersection of a line with
parameter $z$ and column with parameter $w$, etc. With this parameterization, the model has an infinite
family of commuting row-to-row transfer matrices, and can be exactly solved by Bethe Ansatz techniques.
Using recursion relations of Korepin\cite{KO}, Izergin\cite{IZ} obtained a compact determinantal formula for the partition function
of this 6V-DWBC model, defined as the sum over edge configurations of the product of local vertex weights,
divided by the normalization factor $\prod_{i=1}^n c(z_i,w_i)$ (to make the answer polynomial in the $z$'s and $w$'s).
It reads:
\begin{equation}\label{IK} 
Z_{6V}^{(n)}(q;\{z_i\},\{w_j\})=\frac{\prod_{i,j} a(z_i,w_j)b(z_i,w_j)}{\Delta(z)\Delta(w)}
\det_{1\leq i,j\leq n} \left( \frac{1}{a(z_i,w_j)b(z_i,w_j)}\right)
\end{equation}
where $\Delta(z)=\prod_{1\leq i<j\leq n}(z_i-z_j)$ stands for the Vandermonde determinant of the $z$'s.

In the above bijection between 6V configurations and to ASMs, it is easy to track both quantities 
$N(B)$ and ${\rm Inv}(B)$ in terms of 6V weights.
We find that 
$$N(B)= N_{c_2}=\frac{N_c-n}{2} \qquad {\rm Inv}(B)-N(B)=N_{a_1}=N_{a_2}=\frac{N_{a}}{2}$$
where $N_{a_i}$, $N_{b_i}$, $N_{c_i}$ stand for the total numbers vertex configurations of each type.
The determinant result above can therefore be used to compute the refined partition function for ASMs, which 
counts ASMs with a weight $x/y$ per entry $-1$ and a weight $y$ for each inversion (with these weights, the 7
ASMs of size 3 above have respective weights: $1,y,y,y^3,y^2,y^2,x y$). Setting
\begin{equation}\label{xyab}
x=\left( \frac{c}{b}\right)^2\qquad y=\left( \frac{a}{b} \right)^2 
\end{equation}
we have:
\begin{equation}\label{zasm} 
Z_{ASM}^{(n)}(x,y)=\sum_{ASM\, B} x^{N(B)} y^{{\rm Inv}(B)-N(b)}=b^{-n(n-1)} \, Z_{6V}^{(n)}(a,b,c)
\end{equation}
where $Z_{6V}^{(n)}(a,b,c)$ refers to the homogeneous limit of the partition function \eqref{IK}
of the 6V model in which all $a(z_i,w_j)$
are equal to a, etc. (This is obtained by letting all $z_i\to z$ and all $w_i\to w$, with $a=a(z,w)$, $b=b(z,w)$ and $c=c(z,w)$.)
This and more refinements were worked out in \cite{BDZJ1}. We have the following remarkable result:

\begin{thm}\cite{BDZJ1}\label{asmthm}
The partition function for refined ASM reads:
\begin{equation}\label{asmdet}
Z_{ASM}^{(n)}(x,y)= \det_{0\leq i,j \leq n-1} \left ( (1-\nu){\mathbb I}+\nu G\right) 
\end{equation}
where $\nu$ is any solution to the equation
\begin{equation}\label{nudef} x \nu(1-\nu)=\nu +y(1-\nu)\end{equation}
and the $n\times n$ determinant is the principal minor for the $n$ first rows and columns of the infinite
matrix $ M_{ASM}=(1-\nu){\mathbb I}+\nu G$ whose entries are generated by
\begin{equation}\label{fasm} f_{M_{ASM}}(z,w)=\frac{1-\nu}{1-z w}+\frac {\nu}{1-z x-w -(y-x)z w} 
\end{equation}
\end{thm}

We notice that the second term in the generating function is nothing but that of the transfer matrix for 
1+1D Lorentzian gravity, up to some gauge transformation $z\to z/\sqrt{x}$ and $w\to w\sqrt{x}$,
and upon the identification $x=a^2g^2$ and $y=g^2$.

\subsection{DPP from lattice paths}

Descending plane partitions are arrays of positive integers of the form:
$$\begin{matrix} a_{1,1} & a_{1,2} & a_{1,3} & \cdots & \cdots & a_{1,\mu_1-2} & a_{1,\mu_1-1} & a_{1,\mu_1} \\
 & a_{2,2} & a_{2,3} & \cdots & \cdots & a_{2,\mu_2} & & \\
 & & a_{3,3} & \cdots & a_{3,\mu_3} & & & \\
 & & \ddots & \cdots & & & & \\
 & & & a_{r,r}\cdots a_{r,\mu_r} & & & &
\end{matrix}$$
such that the sequece $\mu_i$ is strictly decreasing $\mu_{i+1}<\mu_i$, and that for $\lambda_i=\mu_i-i+1$,
$\lambda_0=\infty$:
$$  a_{i,j}\geq a_{i,j+1} \qquad a_{i,j}>a_{i+1,j} \qquad \lambda_i<a_{i,i}\leq \lambda_{i-1} $$
for all $i,j$.

The integers $a_{i,j}$ are called parts. A DPP is said to be of order $n$ if $a_{i,j}\leq n$ for all $i,j$. A part
$a_{i,j}$ is said to be special if $a_{i,j}\leq j-i$. By convention, the empty partition is a DPP.
In the following we will compute the partition function
$Z_{DPP}^{(n)}(x,y)$ which is the sum over all DPP of order $n$ with a weight $x$ per special part 
and $y$ per remaining part. Here are the 7 DPP of order 3:
$$\emptyset\qquad \begin{matrix} 2 \end{matrix}\qquad \begin{matrix} 3 \end{matrix} \qquad 
\begin{matrix} 3 & 1 \end{matrix}\qquad 
\begin{matrix} 3 & 2\end{matrix}\qquad \begin{matrix} 3 & 3\end{matrix}\qquad \begin{matrix} 3 & 3\\ & 2\end{matrix}$$
They have respective weights: $1,y,y,xy,y^2,y^2,y^3$, as the only special part is the entry $1$ in the fourth DPP.

\begin{figure}
\centering
\includegraphics[width=10.cm]{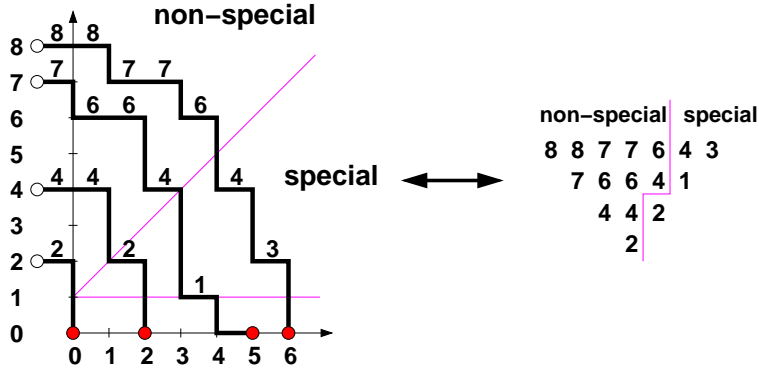}
\caption{\small Non-intersecting lattice path configuration for a sample DPP of order $n\geq 8$.
We have indicated the domains in which horizontal steps correspond to special/non-special parts.}
\label{fig:dppnilp}
\end{figure}

The DPPs are in bijection with configurations of non-intersecting lattice paths illustrated in Fig.\ref{fig:dppnilp}
and defined as follows. We consider the non-negative
quadrant $\Z_+\times \Z_+$ of the integer plane. Paths may take two types of step: vertical up or horizontal left.
They start along the $x$ axis at positions of the form $(s_i,0)$ (recorded from right to left) 
and end along the $y$ axis at positions $(0,s_i+2)$ (recorded from top to bottom)
for $i=1,2,...,r$. We add a final horizontal left step at the end of each path. Reading paths from left to right 
and top to bottom, we record the vertical positions $y=a_{i,j}$ of the $j$-th horizontal step from 
the left taken on the $i$-th path from top (steps with $y=0$ are not recorded). 
These form a DPP with $r$ rows, of order any $n\geq s_1+2$. 
Conversely to each DPP with $r$ rows we may associate such
a path configuration. Note that the starting points are such that $s_i=\lambda_i-1$, where 
$\lambda_i=\mu_i-i+1$ the total number of parts in the row $i$. We note that the special parts correspond to horizontal steps 
taken in the strict upper octant $y\geq x+1$ of the plane, and the remaining parts correspond 
to the horizontal steps in the domain $1\leq y\leq x+1$,
while horizontal steps along the x axis do not count.

The computation of $Z_{DPP}^{(n)}(x,y)$ uses the Lindstr\"om-Gessel-Viennot \cite{LGV1,LGV2} determinant formula
expressing the partition function for non-intersecting lattice paths $Z$ as a determinant $\det(Z_{i,j})$
where $Z_{i,j}$ is the partition function for a single path starting from the $i$-th point and ending at the $j$-th.
This leads to the following:

\begin{thm}\cite{BDZJ1}\label{dppthm}
The partition function $Z_{DPP}^{(n)}(x,y)$ for DPP of order $n$
with weight $x$ per special part and $y$ per other part reads:
\begin{equation}\label{dppdet}
Z_{DPP}^{(n)}(x,y)= \det_{0\leq i,j \leq n-1} \left( {\mathbb I} + H\right)
\end{equation}
where the determinant is the principal minor of size $n\times n$ of the infinite matrix $M_{DPP}={\mathbf I}+H$
with generating function:
\begin{equation}\label{fdpp} f_{M_{DPP}}(z,w)= \frac{1}{1-z w}+\frac{1}{1-z}\frac{y z}{1-x z-w -(y-x) z w}
\end{equation}
\end{thm}

This bears a striking resemblance to the formula of Theorem \ref{asmthm}, and also involves the generating function for the transfer
matrix of 1+1D Lorentzian gravity.

\subsection{Proof of the ASM-DPP conjecture}

The complete MRR conjecture is now proved\cite{BDZJ1}. The proof uses the same ingredients as exposed here,
and includes extra refinements. For pedagogical reasons, we retain here only the special/non-special part dependence
on the DPP side, and the number of $-1$/inversion number dependence on the ASM side. Let us show that:
\begin{equation}\label{ident}
Z_{ASM}^{(n)}(x,y)= Z_{DPP}^{(n)}(x,y) \end{equation}

Both expressions are determinants of the principal minor of size $n$ of some infinite matrix, in other words, these are
the determinants of a finite truncation to the $n$ first rows and columns of infinite matrices. 
There is a very simple relation
(independent of $n$)
between the generating functions of the two infinite matrices $M_{ASM}$ and $M_{DPP}$, namely:
$$(1-z)(1-(1-\nu)w) f_{M_{DPP}}(z,w)=(1-\frac{z}{1-\nu})(1-w)f_{M_{ASM}} (z,w) $$
as a direct consequence of \eqref{nudef}.

Moreover, introducing the strictly lower triangular infinite ``shift" matrix $S$ with entries $S_{i,j}=\delta_{i,j+1}$
for $i,j\geq 0$, we see that for any infinite matrix $A$ and any complex parameters $\alpha,\beta$: 
$(1-\alpha z)f_A(z,w)=f_{({\mathbb I}-\alpha S)A}(z,w)$,
and similarly $(1-\beta w)f_A(z,w)=f_{A({\mathbb I}-\beta S^t)}(z,w)$.
Moreover, as ${\mathbb I}-\alpha S$ is lower uni-triangular and ${\mathbb I}-\beta S^t$ is upper uni-triangular,
any finite truncation to the $n$ first rows and columns of $({\mathbb I}-\alpha S)A$ and 
$A({\mathbb I}-\beta S^t)$ has the same determinant as the corresponding truncation of $A$ alone.
This allows to identify the principal minors of $M_{DPP}$ and $M_{ASM}$ and the desired identity follows.

To conclude, our generating function method has allowed for a simple comparison and eventual proof of the
identity of the partition functions for ASM and DPP. In passing, it shows some mysterious connection to
the quantum integrable transfer matrix of 1+1D Lorentzian gravity.  It seems that the quantum integrability 
of the 6V model is unrelated to this one, and this is yet another puzzling fact about ASMs as a crossroad between
various integrable models.

\section{Discrete Integrable systems and cluster algebras}

\subsection{T-system and initial data}

\begin{figure}
\centering
\includegraphics[width=12.cm]{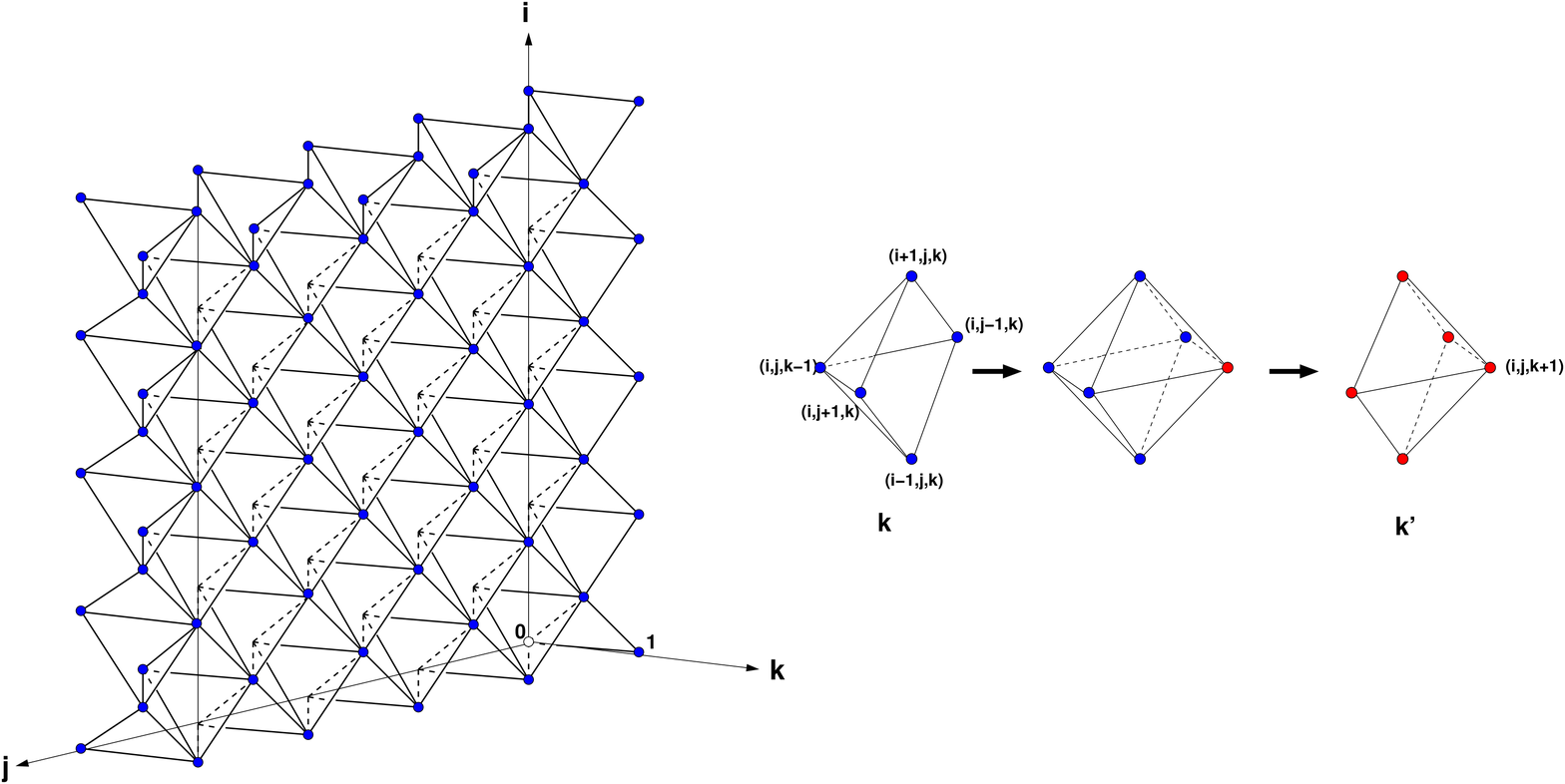}
\caption{\small The ``flat" stepped surface carries all the initial data at times $k=0,1$ for the T-system.
We have represented a local application of the T-system evolution: it evolves the stepped surface by ``adding"
an elementary octahedron to it.}
\label{fig:boundat}
\end{figure}

Let us concentrate on the T-system realtion \eqref{defoT} with for simplicity $\lambda=1$.
We may use this relation to describe the non-linear discrete time evolution ($k\in \Z$) of a 2D space-dependent
quantity (for $i,j\in \Z$), $T_{i,j,k}$, with:
\begin{equation}\label{tmut}
T_{i,j,k+1}=\frac{T_{i,j+1,k}T_{i,j-1,k}+ T_{i+1,j,k}T_{i-1,j,k}}{T_{i,j,k-1}}
\end{equation}
Initial conditions determining $T_{i,j,k}$ entirely consist in assigning values of $T_{i,j,k}$ along any fixed
``stepped surface" ${\mathbf k}=\{(i,j,k_{i,j})\}_{i,j\in \Z}$ with $k_{i,j}\in \Z$
such that $|k_{i,j+1}-k_{i,j}|=|k_{i+1,j}-k_{i,j}|=1$ for all $i,j\in \Z$.
The simplest such surface, denoted by ${\mathbf k}_0$, has $k_{i,j}=i+j$
mod 2, and consists of all the points at times $0$ and $1$ (cf. Fig.\ref{fig:boundat}).
This is the surface used in the definition of the Lambda-determinant.
The evolution \eqref{tmut} may be used to trade a stepped surface $\bf k$ with 
a surface $\bf k'$ identical to it except for the point $(i,j,k_{i,j})\to (i,j,k_{i,j}'=k_{i,j}+2)$. This is however possible only if
the four neighboring points share the same value of $k=k_{i-1,j}=k_{i+1,j}=k_{i,j+1}=k_{i,j-1}=k_{i,j}+1$.
We see that this amounts to completing an octahedron and trading its point with 
the lowest value of $k$ for the one with the largest
(see Fig.\ref{fig:boundat}). The  application of the T-system is sometimes called ``octahedron move".

This will be later interpreted
as a mutation at point $(i,j)$ in the corresponding cluster algebra. 

\subsection{Cluster Algebra}

Here we give a simplified definition of cluster algebra (namely skew-symmetric or geometric type, 
with no coefficient), and we refer to Fomin and Zelevinsky's original paper\cite{FZI} for more general considerations. 

Let $\mathcal T$ be an infinite tree with fixed vertex degree $n$, and edges labeled $1,2,...,n$ around each vertex.
To each vertex $v$ of $\mathcal T$ we attach two data:
(i) a vector of formal variables with $n$ components ${\mathbf x}_v=(x_{1}(v),...,x_{n}(v))$
(ii) an integer skew-symmetric matrix $B_v=(b_{i,j}(v))$, coding a quiver $Q_v$ (graph with oriented edges)
with $n$ vertices labeled $1,2,...,n$, and with no oriented loops of length $\leq 2$.

The data at adjacent vertices $v,v'$ of $\mathcal T$ connected via an edge labeled $k$ are related via a mutation $\mu_k$.
The mutation acts both on quivers and data vectors.

The application of $\mu_k$ on $Q_v$ transforms it as follows: (i) it reflects all
edges of $Q_v$ adjacent to the vertex $k$ (ii) for any path of length $2$ of the form $i\to k\to j$ in $Q_v$,
it adds an oriented edge $i\to j$ to $Q_{P\mu_k(v)}$ unless some edges $j\to i$ already exist, in which case
one of them is canceled. As an example, we represent here the mutations $\mu_1$ and then $\mu_2$
on a quiver $Q$:
$$ \raisebox{-1.cm}{\hbox{\epsfxsize=9.cm \epsfbox{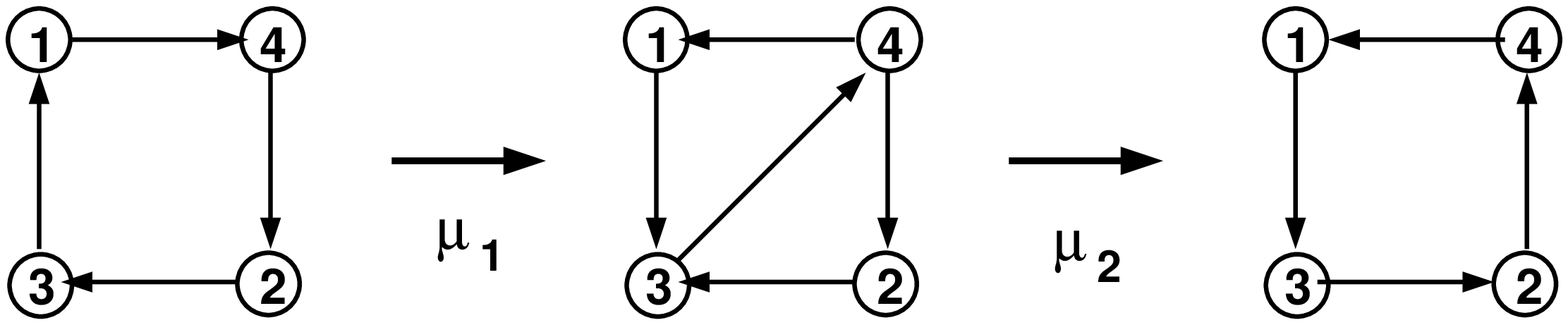}}} 
$$

The application of $\mu_k$ on ${\mathbf x}_v$ leaves $x_i(v)=x_i(v')$ invariant, except for $i=k$,
where 
\begin{equation}\label{mutarul} x_k(v)x_k(v')=\prod_{i\to k\, {\rm in }\, Q_v} x_i(v) +\prod_{k\to j\, {\rm in }\, Q_v} x_j(v)
\end{equation}
In other words, the quiver $Q_v$ at $v$ encodes the transformation of the data vector from $v\to v'$
any adjacent vertex on $\mathcal T$.

The cluster algebra is the algebra with a unit generated by all the cluster variables at all the clusters.
It is entirely determined by the ``initial data" at any fixed vertex of the tree $\mathcal T$.
The data vectors are called clusters, their entries cluster variables. The matrix $B$ is called exchange matrix.
The pair $({\mathbf x},B)$ is called a seed.
The integer $n$ is called the rank of the cluster algebra, and can be infinite.

\subsection{Properties, applications, conjectures}

The major built-in property of cluster algebras is the Laurent phenomenon:
the cluster variables at any vertex of $\mathcal T$ may be expressed as Laurent polynomials of the cluster
variables at any other vertex of the tree. This is not obvious, as the mutations involve many divisions.
The fact that polynomials are always divisible by the new denominators is a
non-trivial  outcome of the definitions.

Two important results have been obtained regarding the classification of cluster algebras. First the 
cluster algebras with finitely many distinct cluster variables (finite type) are classified\cite{FZII}. In our case,
they correspond to an initial quiver $Q$ that is the Dynkin diagram of a simply-laced Lie algebra 
of $A,D,E$ type, with a choice of orientation of the edges. Next, the cluster algebras with
finitely many distinct quivers (mutation-finite) are also classified\cite{FST}. They are essentially made of 
cluster algebras attached to (decorated) triangulations of Teichm\"uller space, plus a few exceptions.

Applications of cluster algebra are very diverse both in mathematics and physics: 
canonical and dual canonical bases of quantum groups,
total positivity, preprojective algebra representations, quiver representations, triangulated 2-Calabi Yau categories,
discrete integrable systems and somos-like sequences, dimer models,
Donaldson-Thomas invariants of topological string theory,
wall-crossing in super-Yang-Mills theories, quantum dilogarithm, Teichm\"uller space geometry, etc.

A mysterious property was conjectured by Fomin and Zelevinsky\cite{FZI}, and was only proved in particular cases so far:
the Laurent phenomenon of cluster algebra is positive, namely the Laurent polynomials involved have only
non-negative integer coefficients. This calls for some combinatorial interpretation, still missing to this day.

From now on, we concentrate on the connection to discrete integrable systems, and more precisely to the T-system.

\subsection{T-system as a sub-cluster algebra}

We have the following:

\begin{thm}\cite{DFK09}
The T-system relation is a mutation in an infinite rank cluster algebra.
\end{thm}

The initial seed is constructed as follows.
The cluster is made of the initial data along the stepped surface ${\mathbf k}_0$ discussed above,
namely with ${\mathbf x}=(x_{i,j})_{i,j\in \Z}$, and $x_{i,j}=T_{i,j,i+j\, {\rm mod}\, 2}$.
The quiver may be represented as an infinite square lattice with vertices $(i,j)\in \Z^2$ 
and oriented edges $(i,j)\to (i,j\pm 1)$ if $i+j=0$ mod 2, and $(i,j)\to (i\pm 1,j)$ otherwise.

Starting from the initial data stepped surface ${\mathbf k}_0$, we may apply a mutation $\mu_{i,j}$ whenever
the octahedron move is permitted at $(i,j,k_{i,j})$ (this limits the number of possible mutations, and therefore we
do not explore the entire cluster algebra in this way). It is easy to check that in all cases the quiver has 2 incoming
and two outgoing edges at each mutable vertex, and the rules \eqref{mutarul} are equivalent to the
T-system relation \eqref{tmut}. (This can be inferred from the example of mutation given above.).

\subsection{Solution and Laurent positivity}

Let us briefly describe the solution of the T-system in the case of the initial data surface ${\bf k}_0$.
The surface can be decomposed into half-octahedra that can be completed upon mutation. We color
in white and grey the triangular faces of each half-octahedron as follows, and attach to the ``rhombi" formed by pairs 
of triangles sharing a horizontal edge (belonging to the $(j,k)$ plane) the following $2\times2$ matrices $D,U$
according to whether the grey triangle is on the bottom or the top of the rhombus:

\begin{equation}
\label{halfocta}
\raise-1.5cm\hbox{ $\epsfxsize=2.5cm \epsfbox{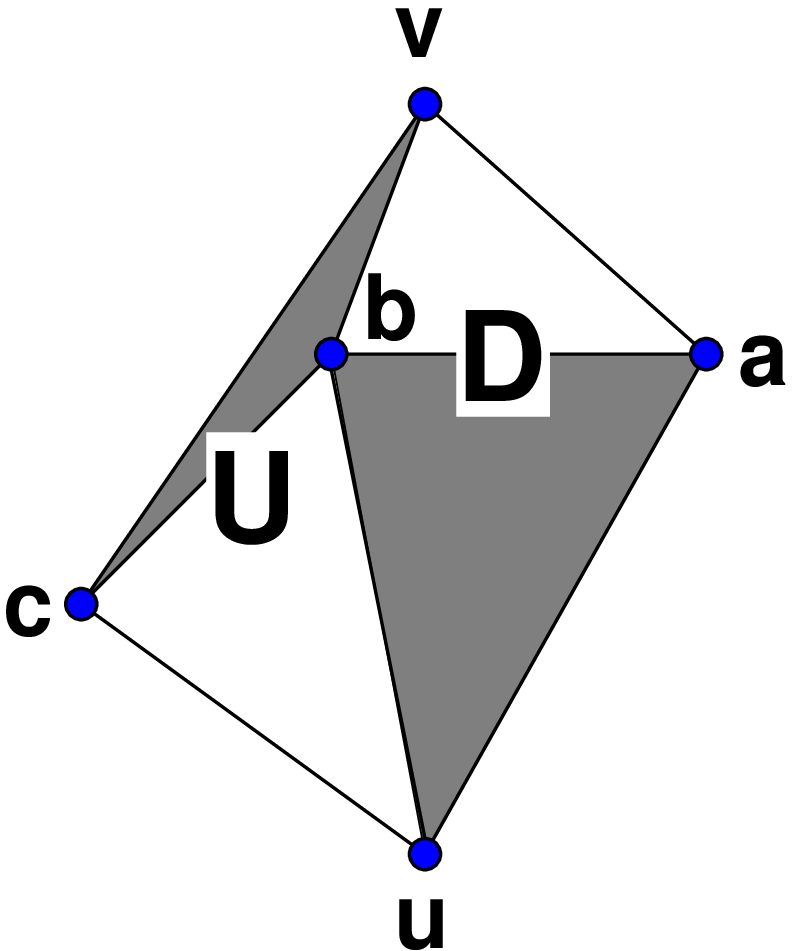}  $} \qquad D_i(u,a,b)=
\begin{pmatrix}
1 & 0\\
\frac{u}{b} & \frac{a}{b}
\end{pmatrix}
\qquad U_i(b,c,v)=
\begin{pmatrix}
\frac{b}{c} & \frac{v}{c}\\
0& 1
\end{pmatrix}
\end{equation}
where $a=T_{i,j-1,k}$, $b=T_{i,j,k-1}$, $c=T_{i,j+1,k}$ $u=T_{i-1,j,k}$ and $v=T_{i+1,j,k}$ and where the subscript $i$ 
indicates that we should think of these $2\times 2$ matrices as embedded into an infinite identity matrix with rows and 
columns indexed by $\Z$, except at rows and columns $i,i+1$ where we insert the $2\times 2$ matrix.

The octahedron move corresponds to the matrix identity
\begin{equation}
\label{nett}
D_i(u,a,b)U_i(b,c,v)=\raise-1.5cm\hbox{ $\epsfxsize=5cm \epsfbox{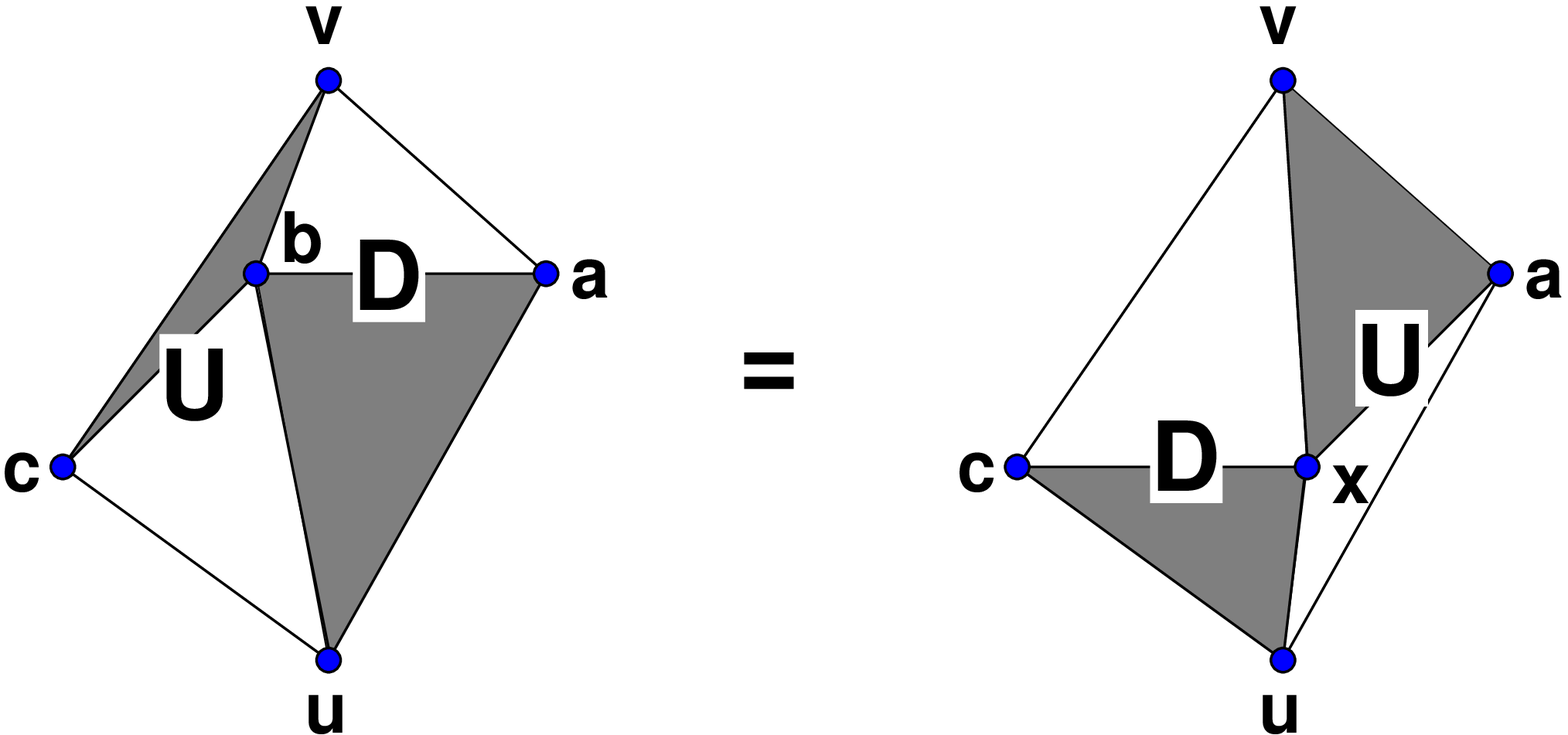}  $}
=U_i(a,x,v)D_i(u,x,c)
\end{equation}
which is satisfied iff
$$ b x=ac+uv $$
By this construction, we represent the T-system mutation by a reordering relation of $D$ and $U$ matrices.

The solution $T_{i,j,k}$ for $k>0$ and $i+j+k=0$ of the T-system with inital data along ${\mathbf k}_0$ can be expressed
explicitly in terms of the assigned values as the determinant of a matrix $M_{i,j,k}$ built out of products of $D$'s and $U$'s. 
The 
product is best expressed pictorially:
$$ M_{i,j,k} = \raise-2.cm\hbox{ $\epsfxsize=4.cm \epsfbox{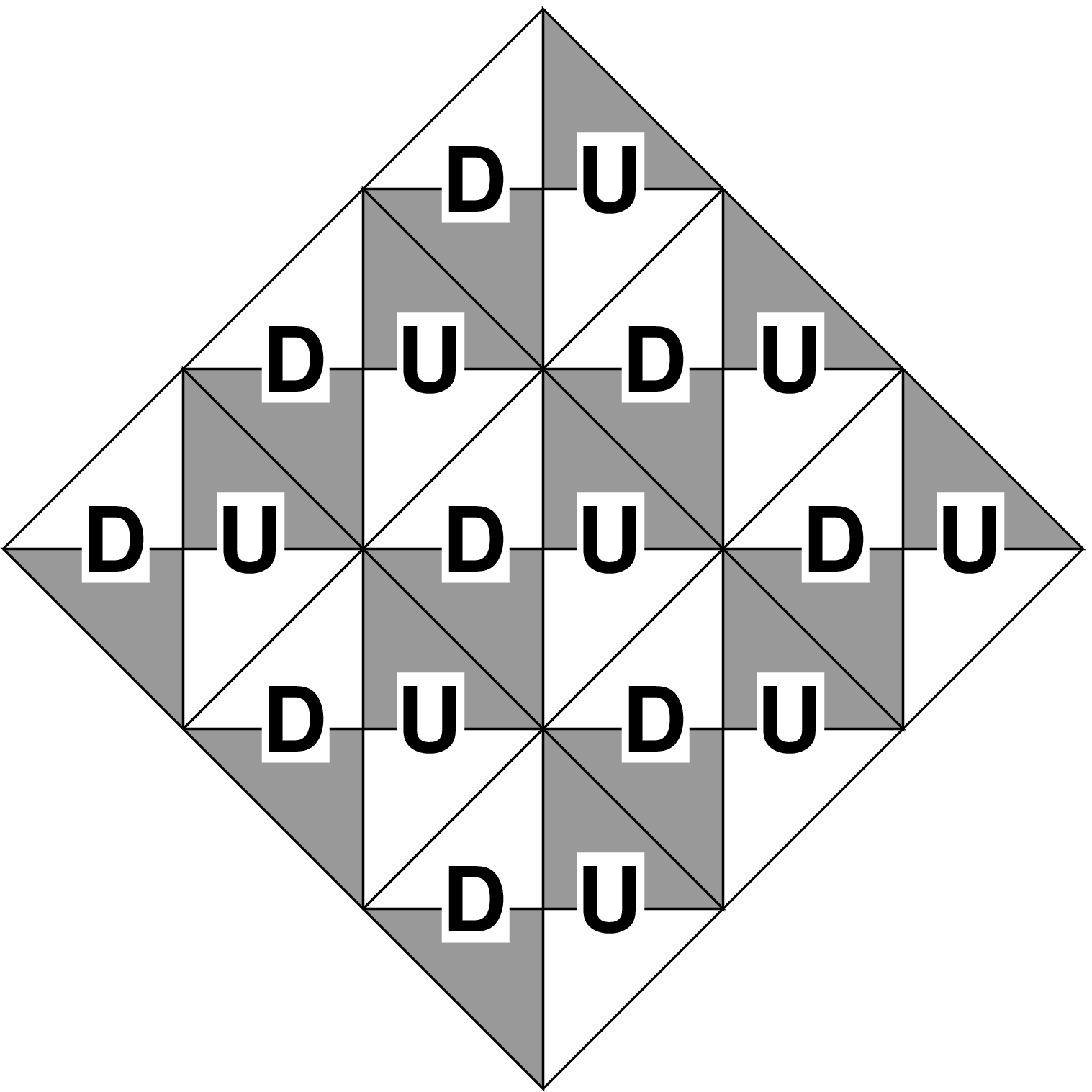}  $} =D_i U_i D_{i-1}U_{i-1}D_{i+1}U_{i+1} ... D_i U_i$$
where the picture is a view from behind of the stepped surface ${\mathbf k}_0$ ( from a point $(i,j,k')$ with $k'<<0$),
the central vertex is $(i,j,i+j\, {\rm mod}\, 2)$ and the size of the tilted square is $k-1$, so that all the boundary vertices
sit at time $k=1$. Moreover, the product is read from left to right, and a D,U is to the left of a D,U in the product iff
it is on the picture, with the obvious fact that $D_i,U_i$ commute with $D_j,U_j$ as soon as $|i-j|>1$. Note that $M_{i,j,k}$
is the embedding of a $2k-2\times 2k-2$ matrix into an infinite identity matrix. Truncating $(M_{i,j,k})_{a,b}$ 
to the square window $a,b\in [i-k+2,i+k-1]$ we may take its determinant, still denoted $\det(M_{i,j,k})$.

We have the final:
\begin{thm}\cite{DFK12}\label{solT}
The solution $T_{i,j,k}$ of the T-system is expressed in terms of the initial data along the stepped surface ${\mathbf k}_0$ as:
$$T_{i,j,k}=\det(M_{i,j,k}) \, \prod_{a=1}^{k-1} T_{i-k+1+a,j-a,1}^{-1}T_{i-k+1+a,j+a,1}$$
\end{thm}

This provides an explicitly positive Laurent polynomial of the initial data, due to the structure of the $D,U$ matrices.
The proof is by induction under octahedron move, making extensive use of the relation \eqref{nett}. This was extended to
arbitrary stepped surface initial data\cite{DFK12} as well as other boundary conditions such as wall $i=$const.
or $j=$const. along which $T_{i,j,k}$ is fixed to be $1$. In all cases positivity follows from explicit
expressions involving the $D,U$ matrices.

\section{Conclusion}

In these notes we have explored various combinatorial problems, with one or more links to integrable systems.
These links seem to always point to the existence of some kind of tree or path model underlying the
original combinatorial problem. Note that a path model is typically a discrete version
of a quantum path integral.

In the last section, the solution of the T-system was expressed in terms of $D,U$ matrices. These can be interpreted as
elementary pieces allowing to construct networks, namely directed graphs with edge weights. More precisely,
we attach to $D_i,U_i$ the following elementary pieces of graph, connecting entry and exit vertices labeled $i,i+1$
via edges oriented from left to right, and whose weights match the corresponding matrix entries:
$$ D_i(u,a,b)= \raise-.7cm\hbox{ $\epsfxsize=2.5cm \epsfbox{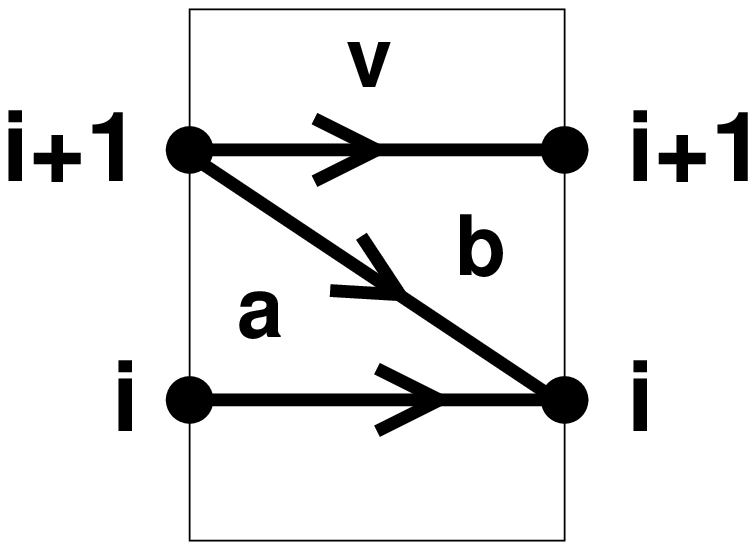}  $}\qquad 
U_i(u,a,b)= \raise-.7cm\hbox{ $\epsfxsize=2.5cm \epsfbox{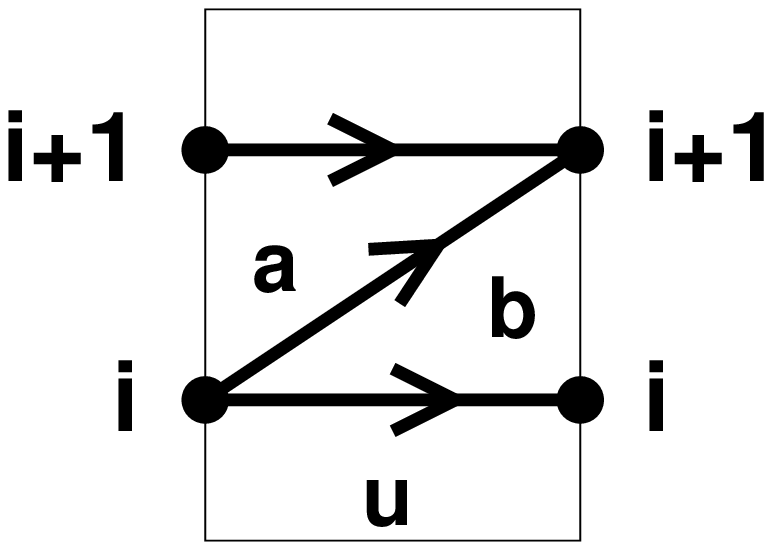}  $}$$
The above products $M_{i,j,k}$ of $D,U$ matrices correspond to the following networks:
$$ \raisebox{-1.cm}{\hbox{\epsfxsize=10.cm \epsfbox{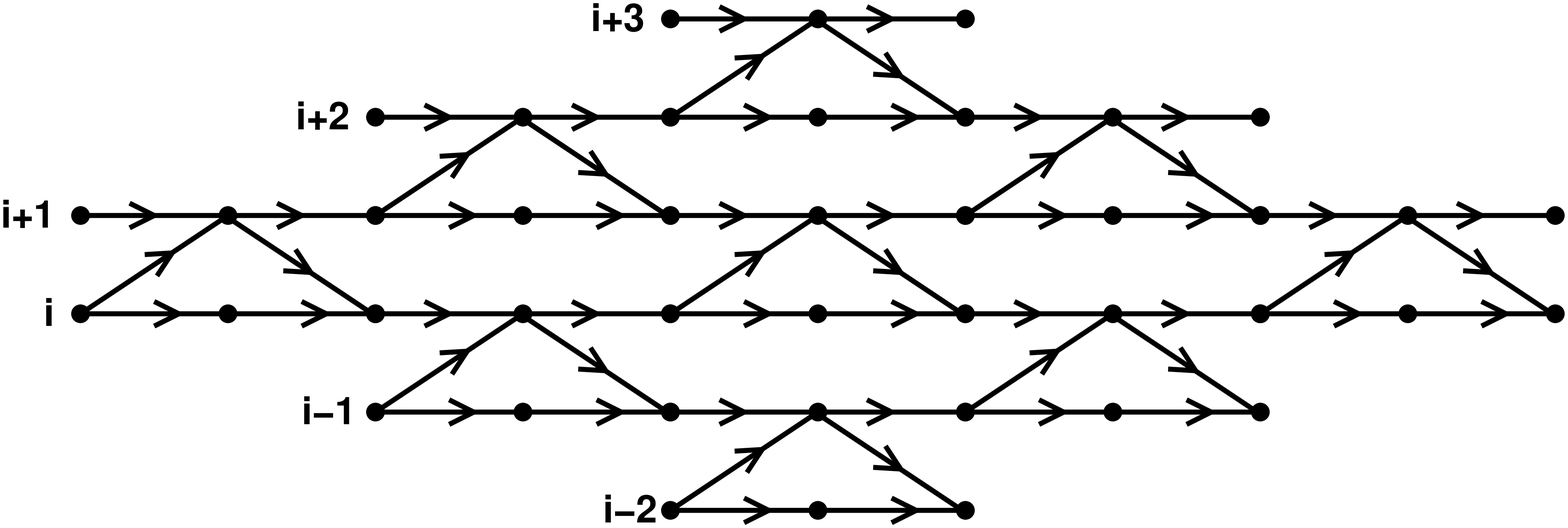}}} $$
with a total of $2k-2$ rows.
By the Lindstr\"om-Gessel-Viennot theorem,
the factor $\det(M_{i,j,k})$ in Theorem \ref{solT}
is the partition function for $k$ non-intersecting paths starting at the $k$ south-west entry points of
the network, and ending at the $k$ south-east exit points of the network. This gives yet another path model for the
solution of the T-system. The positivity of the path weights is responsible for the positive
Laurent phenomenon.

Path models are also very important for non-commutative generalizations. Indeed, a path is a fundamentally
non-commuting object, as its steps are taken in a particular order. Attaching non-commuting weights to paths
yields natural non-commutative expressions. Cluster algebras have a ``quantum" version, in which the cluster
variables have commutation relations encoded by the quiver\cite{BZ}. In rank 2, a fully non-commutative version of cluster algebra
was introduced via path model solutions\cite{DFK09b,DFK10}. Finally, a higher rank non-commutative discrete 
Hirota equation was derived\cite{DFK10} from non-commutative path and continued fraction expressions involving
quasi-determinants of non-commuting variables.


\end{document}